\documentclass[12pt]{iopart}

\usepackage{iopams,graphicx}
\begin{document}

\title{Experimental evidence for an intermediate phase in the multiferroic YMnO$_{3}$}

\author{G. N\'{e}nert$^{1}$, M. Pollet$^{2}$, S. Marinel$^{3}$, G. R. Blake$^{1}$, A. Meetsma$^{1}$ and T. T. M. Palstra$^{1}$}

\address{$^{1}$Solid State Chemistry Laboratory, Zernike Institute for Advanced Materials,\\ University of Groningen, Nijenborg 4, 9747 AG Groningen, The Netherlands}
\address{$^{2}$ICMCB - CNRS Physico-Chimie des Oxides Conducteurs, 87 Avenue du Docteur Schweitzer 33608 Pessac Cedex France}
\address{$^{3}$Laboratoire CRISMAT, UMR CNRS ENSI Caen - 14050 Caen Cedex 4 France}

\begin{abstract}

We have studied YMnO$_{3}$ by high-temperature synchrotron X-ray
powder diffraction, and have carried out differential thermal
analysis and dilatometry on a single crystal sample. These
experiments show two phase transitions at about 1100K and 1350K,
respectively. This demonstrates the existence of an intermediate
phase between the room temperature ferroelectric and the high
temperature centrosymmetric phase. This study identifies for the
first time the different high-temperature phase transitions in
YMnO$_{3}$.
\end{abstract}


\maketitle

\section{Introduction}
In typical ferroelectrics the creation of an electric dipole
moment involves a charge transfer between empty d-shell metal ions
and occupied 2p orbitals of oxygen. This mechanism of
ferroelectricity precludes magnetic moments and magnetic order.
YMnO$_{3}$ is part of a class of materials that exhibit both
electric and magnetic orders, called multiferroics. The
coexistence of ferroelectricity and magnetic order allows the
manipulation of electric and magnetic moments by magnetic and
electric fields, respectively \cite{divers}. In multiferroics, the
charge transfer from the 2p orbitals towards the transition metal
ion is not relevant due to the partially filled nature of the
d-shell. Indeed, in YMnO$_{3}$ the Mn-ion remains in the
barycenter of its oxygen coordination. However, this charge
transfer is allowed towards the empty d-shell of the rare-earth
and creates two dipoles of opposite sign resulting in a
ferrielectric state \cite{ferrielectric}. Recent \emph{ab-initio}
calculations indicate that the ferrielectric state involves no
charge-transfer, but is related to the tilting of the MnO$_{5}$
polyhedra \cite{Aken1}. However, there is no clear
high-temperature structural information about this transition to
clarify the origin of the ferrielectric state in this material.
While a lot of effort is put in the search and design of new
multiferroics, the nature of the mechanism of ferrielectricity and
thus the nature of the coupling between the different degrees of
freedom is still not well understood in YMnO$_{3}$. We can
distinguish two classes of antiferromagnetic ferroelectrics:
T$_{N}$ (N\'{e}el temperature) $>$ T$_{FE}$ (critical temperature
for ferroelectricity) and T$_{N}$ $<$ T$_{FE}$. For the first
class, with compounds like RMn$_{2}$O$_{5}$ and o-RMnO$_{3}$
(orthorhombic), the mechanism of ferroelectricity can be
understood phenomenologically by the existence of a gradient of
the magnetization (incommensurate helicoidal spin structure)
leading to a transverse polarization \cite{Mostovoy, Katsura}. The
coexistence and strong coupling between ferroelectricity and
incommensurate magnetism in o-RMnO$_{3}$ is related to
Dzyaloshinskii-Moriya interactions \cite{Sergienko}. For the
second class, which includes h-RMnO$_{3}$ (hexagonal), the origin
of ferroelectricity is not well understood since the Mn
off-centering is not responsible for the ferroelectricity
\cite{Aken1}. Moreover, the h-RMnO$_{3}$ is particularly
interesting due the high polarization at room temperature
(P$_{S}$$\sim$5.5 $\mu$C/cm$^{2}$). The h-RMnO$_{3}$ exhibit
T$_{FE}$$\sim$1000 K and T$_{N}$$\sim$100 K. Despite this
difference, recent studies of h-RMnO$_{3}$ show that the
ferroelectric and antiferromagnetic order parameters are coupled
\cite{Huang,Iliev,Fiebig,Lorentz}.

YMnO$_3$ adopts the centrosymmetric space group $P6_3$/$mmc$ in
its high temperature paraelectric state and the polar space group
$P6_3cm$ at room temperature with a unit cell that is tripled in
volume (the $a$ and $b$ axes are lengthened by a factor of
$\sqrt{3}$). However, the ferroelectric transition temperatures
reported in the literature fall into two distinct groups, and when
taken together, previous studies of YMnO$_3$ suggest that there
are not one but two distinct phase transitions above room
temperature \cite{review}. However, these phase transitions have
never both been observed in a single experiment. Moreover it has
consistently been reported that the ferroelectric transition of
YMnO$_{3}$ occurs at 933K, and not higher as claimed by Abrahams
\emph{et al.} \cite{Abrahams1}. If an Intermediate Phase (IP)
exists, its symmetry and nature have not yet been determined.

In recent years, two structural studies have been carried out on
h-RMnO$_{3}$ powder in the high  temperature regime either by
synchrotron or by neutron diffraction \cite{Katsufuji,Lonkai}.
Both experiments failed to observe any structural phase
transitions even though measurements were performed up to 1000K
for R = Y (synchrotron), up to 1300K for R = Lu, 1100K for R = Yb
and 1400K for R = Tm (neutron). Lonkai \emph{et al.} realized
using symmetry arguments that two high temperature phase
transitions are necessary \cite{Lonkai}. Therefore, they proposed
an antiferroelectric IP having the same symmetry as the room
temperature phase (P6$_{3}$cm). We believe that the absence of
observations of phase transitions in these two recent studies,
contradicting older results, has caused confusion regarding the
high temperature behavior of YMnO$_{3}$. The succession of three
states has thus far been only theoretically incorporated in
investigations of the origin of the ferroelectric state
\cite{Aken1,Fennie}. The aim of this paper is to prove
experimentally the existence of an IP in YMnO$_{3}$. Furthermore,
we will show that our data are in agreement with previous
experimental observations.

First, we report the room temperature structure of a high quality
single crystal. Using dilatometry and differential thermal
analysis, we demonstrate in a single experiment the existence of
an intermediate phase stable between T$_{C1}$ and T$_{C2}$. In
addition, based on synchrotron X-ray powder diffraction, we
confirm the existence of the paraelectric phase above T$_{C2}$
with the P6$_{3}$/mmc symmetry. Finally, we discuss the
differences in transition temperatures between our results and
earlier reports in the light of single crystal quality.

\section{Experimental Techniques}

YMnO$_{3}$ powder was synthesized by reacting stoichiometric
amounts of Y$_{2}$O$_{3}$ (4N) and MnO$_{2}$ (3N metal basis) in
nitrogen atmosphere at 1200$^{\circ}$C for 10h. The powder was
reground and resintered once under the same conditions to improve
crystallinity. A single crystal was grown from the powder in air
by the Floating Zone technique using a four mirror furnace. The
good crystallinity of the crystal was confirmed by Laue
diffraction.

We have carried out an additional high temperature synchrotron
X-ray experiment on the beamline ID31 at ESRF (Grenoble, France)
using a wavelength of 0.3 \r{A}. The powder sample was placed in a
spinning Pt capillary which was heated by means of a three mirror
furnace. The temperature control was monitored by the position of
the Pt peaks with a feedback system to compensate for the aging of
the lamps. Rietveld refinements of the data were carried out using
the GSAS software package \cite{Larson}.

In order to check the quality of our single crystal, we carried
out a full data set collection at room temperature. A black
crystal with approximate dimensions of
0.14$\times$0.12$\times$0.04 mm was measured at room temperature
on a Bruker SMART APEX CCD diffractometer. Intensity measurements
were performed using graphite monochromatic Mo-K$_{\alpha}$
radiation. The final unit cell was obtained from the xyz centroids
of 1743 reflections after integration. Intensity data were
corrected for Lorentz and polarization effects and for absorption
\cite{Sadabs}. The program suite SHELXTL was used for space group
determination assuming an inversion twin \cite{XPREP}.

The differential thermal analysis (DTA) was carried out in a flow
of nitrogen using a SDT2960 of TA Instruments between room
temperature and 1473K with a ramp of 5K/min. We used a crushed
single crystal for this experiment in a platinum pan.

The dilatometry measurements were carried out in static air using
a SETARAM TMA92 dilatometer. The temperature ramp was 1 K/min
starting from room temperature to 1473K. A load of 10 g was
applied on the single crystal to optimize the signal to noise
ratio. The choice of this load was based on several trial runs.
The sample used was cut in a rectangular shape, but not oriented
along a particular crystallographic orientation. The single
crystal was protected from the alumina parts using platinum
sheets. All measurements were corrected for the blank signal
(alumina + platinum).

\section{Results-Discussion}

The refined atomic coordinates and isotropic displacement
parameters of our crystal at room temperature are presented in
Table \ref{table1}. The crystal contains inversion twins as a
natural consequence of the polar structure. The structure is very
similar to that previously reported \cite{Katsufuji}. However,
there is a significant difference in the unit cell volume of our
crystal compared with the structures of the single crystals grown
from Bi$_{2}$O$_{3}$ flux (see Table 2), even when experimental
uncertainty is taken into account. We will discuss later the
possible origin of such difference and its consequence for the
ferroelectric transition temperature.

\begin{table}[htb]
\centering
\begin{tabular}{|c|c|c|c|c|}
\hline
Atoms & x & y & z & U$_{eq}$ (\r{A}$^{2}$) \\
\hline
Y$_{1}$  &      0&   0 &  0.7711(2)  &     0.0056(5)\\
\hline
Y$_{2}$  &     1/3&  2/3 &  0.73038(-) &     0.0060(3) \\
\hline
Mn       &     0.6672(5) &  0  & 0.4976(2)  &     0.0059(5)\\
\hline
O$_{1}$  &     0.697(1)&  0 &  0.661(1) &     0.010(2)\\
\hline
O$_{2}$  &      0.353(1)& 0 &  8352(9)  &     0.005(2)\\
\hline
O$_{3}$  &      0&      0 &  0.973(2)   &     0.008(3)\\
\hline
O$_{4}$  &     1/3&     2/3 &  0.515(1) &     0.010(3)\\
\hline
\end{tabular}
\\
\caption{Atomic coordinates of a single crystal YMnO$_{3}$ grown
by the Floating Zone Technique from data collected at room
temperature. The cell parameters are \emph{a} = 6.1469(6)\r{A} and
\emph{c} = 11.437(1)\r{A} (space group P6$_{3}$cm, wR(F$^{2}$) =
8.45\% and R(F) = 3.43\%). In parentheses, we indicate the
standard uncertainty. The z coordinate of Y$_{2}$ was fixed due to
the lack of a center of symmetry.}\label{table1}
\end{table}

\begin{table}[htb]
\centering
\begin{tabular}{|p{2cm}|p{2cm}|p{2cm}|}
\hline
Reference& \emph{a} (\r{A}) & \emph{c} (\r{A})\\
\hline
Ref. 19 & 6.145 & 11.42\\
\hline
Ref. 20 & 6.1387(3) & 11.4071(3)\\
\hline
Ref. 21 & 6.125(1) & 11.41(1)\\
\hline
This work & 6.1469(6) & 11.437(1)\\
\hline
\end{tabular}
\\
\caption{Comparison of the lattice parameters between single
crystals grown from a Bi$_{2}$O$_{3}$ flux and this
work.}\label{table2}
\end{table}

\begin{figure}
\centering
\includegraphics[width=8cm]{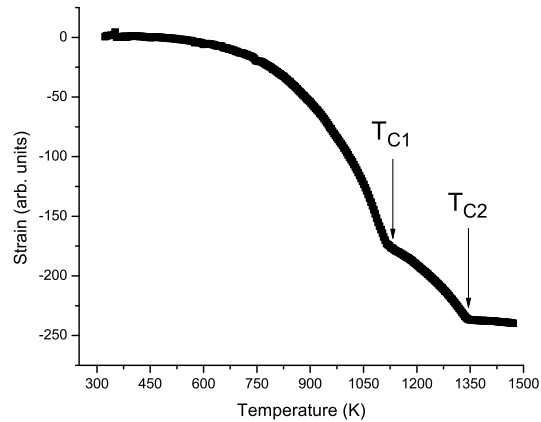}\\
\caption{Dilatometry measurement of an unoriented single crystal
of YMnO$_{3}$. The arrows indicate the two transition temperatures
$_{C1}$ and T$_{C2}$.}\label{fig2}
\end{figure}

In Fig. \ref{fig2}, we show the temperature dependence of the
dilatometric response of an unoriented single crystal of
YMnO$_{3}$. We observe clearly two anomalies while heating. The
first anomaly, denoted T$_{C1}$, appears at about 1125K and the
second anomaly appears at T$_{C2}$$\sim$1350K. These transitions
are also observed in our DTA experiment, shown in Fig. \ref{fig3},
where we notice anomalies at about 1100K for T$_{C1}$ and at
$\sim$1350K for T$_{C2}$.

\begin{figure}[htb]
\centering
\includegraphics[width=8cm]{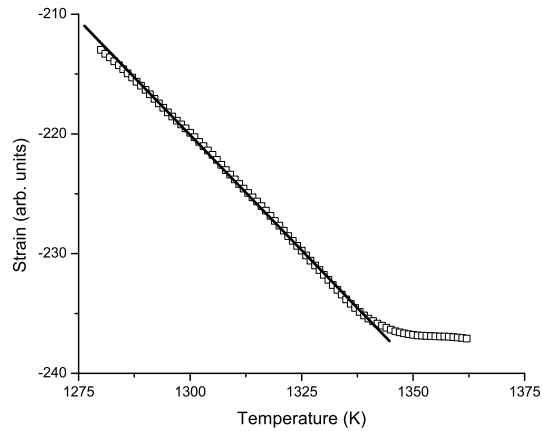}\\
\caption{Linear temperature dependence below T$_{C2}$ of the
thermomechanical analysis.} \label{TMA-detail}
\end{figure}

\begin{figure}
\centering
\includegraphics[width=8cm]{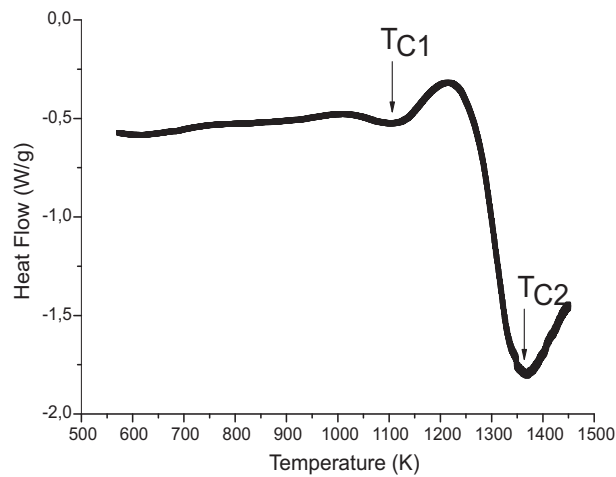}\\
\caption{Differential Thermal Analysis of a crushed single crystal
of YMnO$_{3}$. The measurement was carried out in a N$_{2}$
atmosphere with a rate of 5K/min.The arrows indicate the
ferroelectric transition temperature T$_{C1}$ and the tripling of
the unit cell T$_{C2}$.}\label{fig3}
\end{figure}

We had a closer look at the temperature dependence of the
thermomechanical analysis (TMA) response close to T$_{C2}$. In
Fig. \ref{TMA-detail}, we show a zoom of the TMA response close to
T$_{C2}$, which is linear below about 1340K. While measuring the
TMA response, one measures the strain which usually plays the role
of a secondary order parameter during a phase transition. Indeed,
we know that the transition from the HT phase to the IP could
involve some strain as a secondary order parameter
\cite{toledano}. The linear dependence on temperature of the TMA
response suggests that the transition at T$_{C2}$ is of 2$^{nd}$
order.

In order to clarify the nature of the two phase transitions, we
carried out a DTA experiment on a crushed single crystal under
N$_{2}$ atmosphere. The results are presented in Fig. \ref{fig3}.
We notice a small anomaly at T$_{C1}$ and a larger one at
T$_{C2}$. The differentiation between 2$^{nd}$ order and 1$^{st}$
order phase transitions using DTA remains a non-straightforward
task \cite{DTA2}. While we observe a large peak at about 1350K,
the DTA anomaly at about 1100K is small . This suggests that this
transition at 1350K requires significantly more energy than the
one from IP towards the RT phase

\begin{figure}[htb]
\centering
\includegraphics[width=10cm]{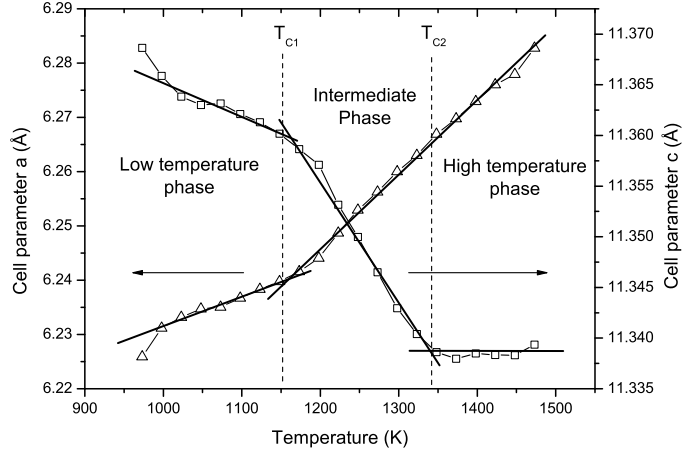}\\
\caption{Cell parameters obtained from a powder sample at the ID31
beamline at ESRF with P6$_{3}$cm symmetry. The high temperature
data above T$_{C2}$ have the symmetry P6$_{3}$/mmc and its cell
parameters are \emph{a}'=$\frac{\emph{a}}{\sqrt{3}}$ and \emph{c}'
= \emph{c}.} \label{cell-parameters-ESRF}
\end{figure}

We further investigated YMnO$_{3}$ using synchrotron X-ray powder
diffraction. We collected data at several temperatures,
concentrating on the range 950K-1475K. This is below and above the
transition temperature T$_{C2}$ observed in the dilatometry and
DTA experiments. We present the measured cell parameters in Fig.
\ref{cell-parameters-ESRF}.

\begin{figure}[htb]
\centering
\includegraphics[width=16cm]{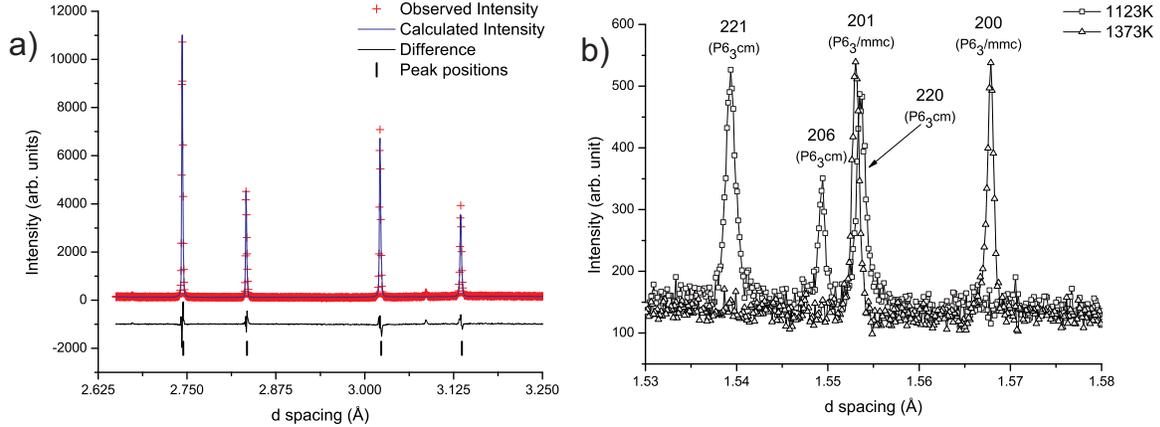}\\
\caption{a) Portion of the Rietveld refinement using the
P6$_{3}$/mmc symmetry at 1373K, b) Disappearance of the (206)
reflection as function of temperature determined by X-ray
synchrotron diffraction. This evidences the transition towards the
P6$_{3}$/mmc symmetry.} \label{Tripling-P63mmc}
\end{figure}

In Fig. \ref{cell-parameters-ESRF}, we can clearly see an anomaly
around 1350K characterizing the transition towards the
high-temperature paraelectric phase with P6$_{3}$/mmc symmetry
\cite{Lonkai,Lukaszewicz}. The transition at T$_{C2}$ is confirmed
by the disappearance of the (206) reflection, which is
characteristic of the tripled unit cell, as shown in Fig.
\ref{Tripling-P63mmc}.

We summarize in Table \ref{table-ESRF-P63mmc} the atomic
coordinates in the high temperature centrosymmetric phase
P6$_{3}$/mmc at 1373K. A physically meaningful model could be
reached only by constraining all the U$_{iso}$ of the different
atoms to be equal.

\begin{table}[htb]
\centering
\begin{tabular}{|c|c|c|c|c|}
\hline
Atom & x & y & z & U$_{iso}$\\
\hline
Y       &  0  &  0  &   0   & 0.0270(7)\\
\hline
Mn      & 1/3 & 2/3 &  1/4 & 0.0270(7)\\
\hline
O$_{1}$ &  0  &  0  &  1/4 & 0.0270(7)\\
\hline
O$_{2}$ & 1/3 & 2/3 &  0.0821(8) & 0.0270(7)\\
\hline
\end{tabular}
\\
\caption{Atomic coordinates determined by X-ray synchrotron
Rietveld refinement at 1373K in the P6$_{3}$/mmc symmetry. Cell
parameters: \emph{a} = 3.62178(2) \r{A} and \emph{c} = 11.3379(2)
\r{A}.}\label{table-ESRF-P63mmc}
\end{table}

We observe that the transition at T$_{C2}$ in Fig.
\ref{cell-parameters-ESRF} does not show any discontinuity. Thus
in the light of the results of dilatometry data, DTA and
diffraction data, we interpret both transitions at T$_{C1}$ and at
T$_{C2}$ as second order phase transitions. We notice that the
transition at T$_{C1}$ is more obvious in the dilatometry
experiment than in the synchrotron X-ray powder data. We observe
that the transition at T$_{C1}$ in the powder sample appears to be
broadened. This type of phenomenon is often seen in
polycrystalline samples where different grains will undergo the
transition at slightly different temperatures due to the effects
of factors such as grain size, morphology and strain. Indeed a
change in slope of the \emph{c} cell parameter is seen in the
vicinity of T$_{C1}$ (see Fig. \ref{cell-parameters-ESRF}).

As we have already remarked, the unit cell for single crystals
grown from a Bi$_{2}$O$_{3}$ flux is systematically smaller than
for crystals grown by the Floating Zone technique. There are two
possibilities to explain this difference: a significant amount of
Bi could be incorporated in the structure and/or the amount of
vacancies can be different in the crystal structure caused by the
different growth techniques.

We expect that if Bi is incorporated into the structure, it will
preferentially occupy the 7-fold coordinated yttrium site. This
hypothesis can be disregarded for several reasons. First,
Bi$^{3+}$ has a larger Shannon radius than Y$^{3+}$: r(Bi$^{3+}$)
= 1.24\r{A} (derived from the average between 6-fold coordination
and 8-fold coordination) and r(Y$^{3+}$)=1.1\r{A} \cite{Shannon}.
Therefore, we expect an increase of the cell parameters. Second,
we have carried out refinements of the occupation on both Y-sites
using the dataset of van Aken \emph{et al.} \cite{Aken2} and our
data. We did not find any evidence for substitution of Y by Bi.
Thus, we have no experimental evidence for the incorporation of Bi
into the lattice.

The second possible explanation for the smaller unit cell of
flux-grown crystals is the creation of vacancies. According to
Abrahams \cite{Abrahams2}, vacancies would also affect the
ferroelectric transition temperature, which can be expressed as
(in Kelvin):

\begin{equation}
\centering T_{FE}=(\kappa/2k)(\Delta z)^{2}
\end{equation}

where $\kappa$ is a force constant, k is Boltzmann's constant and
$\Delta$z is the largest displacement along the polar c axis
between the ferroelectric phase and the paraelectric phase. This
value is expressed in \r{A} due to
$\Delta$z=(z$_{para}$-z$_{ferro}$)$\times$c. The ratio $\kappa$/2k
has been estimated as 2.00(9)$\times$10$^{4}$ K\r{A}$^{-2}$.
According to equation 1, a decrease of T$_{FE}$ implies a decrease
of the force constant, the displacement of the rare earth
z$_{para}$-z$_{ferro}$ and/or the c lattice parameter. It is
obvious that defects can be responsible for such changes. For
instance, an excess in yttrium (YMn$_{1-x}$O$_{3-3/2x}$) would
certainly both decrease the \emph{c} parameter (consistent with
the observations) and lower $\kappa$. In the same way, impurities
can lower T$_{FE}$. Experiments using alumina crucibles instead of
platinum can easily result in YMn$_{1-x}$Al$_{x}$O$_{3}$. Such a
substitution would again account for a small decrease in the c
parameter. All of these suggestions need to be checked
experimentally. We note that the presence of oxygen vacancies
within the structure is unlikely because they would lead to
electrostatic repulsions between the cations, enlarging the
unit-cell.

The expected dependence of T$_{C1}$ on the concentration of
vacancies can explain the discrepancies between the recent results
of Lonkai and Katsufuji and those in the older literature. Both
groups used powders whereas all previous studies were performed on
single crystals grown from a Bi-flux. Both experiments failed to
observe a phase transition, except for R=Tm where Lonkai \emph{et
al.} \cite{Lonkai} observed the start of a transition, however
without reaching the IP. Katsufuji \emph{et al.} measured until
1000K for YMnO$_{3}$, synthesized in air, while our experiments
show a transition at 1100K. For R=Lu and R=Yb, neither Lonkai
\emph{et al.} \cite{Lonkai} nor Katsufuji \emph{et al.}
\cite{Katsufuji} measured at sufficiently high temperatures to
observe a phase transition. For the smaller rare-earths the
ferroelectric transition is expected to be higher than for R=Y
\cite{Abrahams1}. Therefore, our work seems in agreement with
previous literature, if one accepts that YMnO$_{3}$ grown in a
Bi$_{2}$O$_{3}$ flux exhibits suppressed transition temperatures.

\section{Conclusion}

We provide experimental evidence for two high temperature phase
transitions in YMnO$_{3}$ using dilatometry, DTA and synchrotron
X-ray powder diffraction. This is the first proof in one single
experiment for the existence of an intermediate phase in
YMnO$_{3}$. We describe these transitions at T$_{C1}$ and T$_{C2}$
as being likely second order in nature. Above T$_{C2}$, we confirm
the existence of a paraelectric phase with the P6$_{3}$/mmc
symmetry. We relate the higher transition temperatures of our
crystal to the synthesis technique that we used. However, the
nature of the intermediate phase remains to be investigated.
Further experiments are in progress in order to give a full
description of this system.

\section{Acknowledgement}

We thank M. Brunelli, C. V. Colin, A. J. C. Buurma and N. Mufti
for technical support during the experiment at ESRF. The work was
supported by the Dutch National Science Foundation NWO by the
breedtestrategieprogramma of the Materials Science Center,
MSC$^{+}$. M.P. gratefully acknowledges the French Minist\`{e}re
de la Recherche et de la Technologie and the D\'{e}l\'{e}gation
R\'{e}gionale \`{a} la Recherche et \`{a} la Technologie,
r\'{e}gion Aquitaine, for financial support.

\end{document}